\documentclass[paper,prd,reprint,preprintnumbers,nofootinbib,notitlepage,showpacs,showkeys]{revtex4-2}
\usepackage{latexsym,graphicx,amssymb,amsmath,mathrsfs} 
\usepackage{slashed,mathbbol}
\usepackage[utf8]{inputenc}

\DeclareSymbolFont{bbold}{U}{bbold}{m}{n}
\DeclareSymbolFontAlphabet{\mathbbold}{bbold}

\newcommand{\be}{\begin{equation}}      
\newcommand{\ee}{\end{equation}}      
\newcommand{\bea}{\begin{eqnarray}}      
\newcommand{\eea}{\end{eqnarray}}    
\newcommand{\Tr}{\,\textrm{Tr}\,}
\newcommand{\ife}{\,\textrm{if}\,}
\newcommand{\els}{\,\textrm{else}\,} 
\newcommand{\ns}{\,\textrm{ns}\,} 
\newcommand{\s}{\,\textrm{s}\,} 
\newcommand{\MeV}{\,\textrm{MeV}\,} 
\newcommand{\GeV}{\,\textrm{GeV}\,} 
\newcommand{\en}{\,\textrm{end}\,} 
\renewcommand{\top}{\,\textrm{top}\,} 
\newcommand{\fm}{\,\textrm{fm}\,}

\makeatletter
\renewcommand\appendix{\par
\setcounter{section}{0}%
\setcounter{subsection}{0}%
\gdef\thesection{\appendixname\space\@Alph\c@section}}

\long\def\unmarkedfootnote#1{{\long\def\@makefntext##1{##1}\footnotetext{#1}}}
\makeatother

\begin{document} 

\title{Backreaction of mesonic fluctuations on the axial anomaly at finite temperature} 
\author{G. Fej\H{o}s}
\email{gergely.fejos@ttk.elte.hu}
\author{A. Patk\'os}
\email{patkos@galaxy.elte.hu}
\affiliation{Institute of Physics, E\"otv\"os University, 1117 Budapest, Hungary}

\begin{abstract}
{The impact of mesonic fluctuations on the restoration of the $U_A(1)$ anomaly is investigated nonperturbatively for three flavors at finite temperature in an effective model setting. Using the functional renormalization group, the dressed, fully field-dependent Kobayashi-Maskawa--'t Hooft (KMT) anomaly coupling is computed. It is found that mesonic fluctuations strengthen this signature of the $U_A(1)$ breaking as the temperature increases. On the other hand, when instanton effects are included by parametrizing the explicit temperature dependence of the bare anomaly parameter in consistency with the semiclassical result for the tunneling amplitude, a natural tendency appears diminishing the anomaly at high temperatures. As a result of the two competing effects, the dressed KMT coupling shows a well-defined intermediate strengthening behavior around the chiral (pseudo)transition temperature before the axial anomaly gets fully suppressed at high temperature.  As a consequence, we conclude that below $T \sim 200 \MeV$ the $U_A(1)$ anomaly is unlikely to be effectively restored. Robustness of the conclusions against different assumptions for the temperature dependence of the bare anomaly coefficient is investigated in detail.}
\end{abstract}

\maketitle

\section{Introduction}

Understanding the thermal evolution of the anomalous breaking of the axial $U_A(1)$ subgroup of $U_L(N_f)\times U_R(N_f)$ chiral symmetry of QCD remains unsettled. Clarification of this subject is of particular importance for tracking the finite temperature variation of the $\eta-\eta^\prime$ mass difference and also in axion phenomenology. The most important signatures of the $U_A(1)$ anomaly is the absence of expected mass degeneracies (see below) and a nonzero topological susceptibility, which is related to the fluctuations of the topological charge in the QCD vacuum.

It is now basically textbook material that the $U_A(1)$ symmetry recovers at high enough temperature ($T$), but very little is known for certain around, and especially below the (pseudo)critical temperature, $T_c$, of the chiral restoration. If $T\gg T_c$, then due to Debye screening, the instanton density and thus the topological susceptibility show an exponential damping \cite{schaefer96,schaefer98,fukushima11}, but semiclassical calculations definitely fail if $T \lesssim T_c$.  

Whether the $U_A(1)$ symmetry is recovered at the critical temperature has consequences regarding the order of the transition. For massless quarks with two flavors, if the anomaly remains strong at $T_c$, the order of the phase transition is expected to be of second order [with $O(4)$ critical exponents], while if the anomaly is absent, then it is likely to be driven first order due to fluctuations \cite{pisarski84}.  Note that the possibility of a fluctuation-induced first-order transition is based on the $\epsilon$ expansion of the renormalization group (RG) flows and has been also in doubt in past years \cite{pelissetto13,grahl14,nakayama14}.

Studies on the finite temperature recovery of the $U_A(1)$ symmetry have a huge body of literature \cite{lahiri21}. Most works were done on the lattice for two-flavor QCD, and one usually measures to what extent appropriate masses and susceptibilities ($\chi$) are degenerate as a function of the temperature. Specifically, since the pion (pseudoscalar isotriplet meson, $\pi$) and the $a_0$ (scalar isotriplet meson) are related by a $U_A(1)$ transformation, the $m_{a_0}-m_{\pi}$ or $\chi_{\pi}-\chi_{a_0}$ differences can be seen as a good measure of the anomalous $U_A(1)$ breaking.  

In Refs. \cite{bazazov12,buchoff14,bhattacharya14}, for various pion masses, using domain-wall fermions, based on calculating the susceptibilities, the conclusion was that the anomaly is still visible beyond the critical temperature. In Ref. \cite{brandt16} (for $N_f=2$), however, using Wilson fermions, the measurement of the mass difference showed that $U_A(1)$ symmetry is effectively restored at $T_c$ in the chiral limit.  Analyses of the eigenvalue spectrum of the Dirac operator using overlap fermions showed that the susceptibility difference is nonzero, which indicated that the anomaly is present even beyond the critical temperature \cite{dick15}. More recently, chiral extrapolation regarding the susceptibilities using highly improved staggered quark action led to a broken axial symmetry even at $1.6 T_c$ \cite{ding21}. Similar conclusions are drawn in Ref. \cite{kaczmarek21} just above $T_c$. Ensembles generated by two-flavor (Möbius) domain wall sea quarks, and the respective eigenvalue analysis of the susceptibilities may show, on the contrary, that all results are consistent with the axial anomaly being restored at $T_c$ in the chiral limit \cite{tomiya16,aoki21}. 

The strong $CP$ problem and its possible resolution via the axion field are also closely related to the $U_A(1)$ anomaly. They are also of huge importance in current dark matter research. For comprehensive studies on the axion phenomenology and its connection with the axial anomaly, the reader is referred to Refs. \cite{azcoiti16,bonati16,lombardo20}.

Tackling the problem of the $U_A(1)$ restoration can also be approached via several effective models and methods, e.g., the nonlinear sigma model with (unitarized) chiral perturbation theory \cite{gomeznicola16,gomeznicola21}, the Polyakov quark meson model \cite{rai20,li20},  the Polyakov-loop extended Nambu–Jona-Lasinio model \cite{ishii16,ishii17}, the Witten–Di Vecchia–Veneziano model and the extended linear sigma model \cite{bottaro20}. The behavior of the meson spectra and that of the topological susceptibility were investigated using the Dyson-Schwinger approach in Refs. \cite{horvatic19,horvatic20}, exploiting related Ward identities in Ref. \cite{gomeznicola19}, and also with renormalization group techniques \cite{mitter14,braun20}. 

Previously, mesonic fluctuation effects with regard to the $U_A(1)$ anomaly were explored in terms of the three-flavor linear sigma model \cite{fejos16}, where the degree of the $U_A(1)$ breaking was associated with the strength of the Kobayashi-Maskawa--t' Hooft (KMT) determinant coupling. Contrary to the usually expected scenario, it was found that the KMT coupling may grow as the chiral condensates evaporate. This study was based on a chiral invariant expansion of the effective potential, where coupling constants were promoted to coupling (or coefficient) functions that depend explicitly on the chiral condensates. The aforementioned result was obtained via a rather crude approximation of the functional renormalization group (FRG) flows of the coefficient functions, where the anomaly was treated perturbatively to linear order in all of the loop integrals. That is to say, the scale evolution of the anomalous term of the effective action was functionally linear in itself and its derivatives with respect to the chiral condensates. More importantly, its effects on the chirally symmetric part of the effective potential was completely neglected. One of the main goals of this paper is to investigate whether a similar result as of Ref. \cite{fejos16} can be obtained under a more sophisticated approach, where the anomaly function is treated on equal footing with the chirally symmetric part of the effective potential,  i.e.,  no expansion is performed in mesonic loop integrals in terms of the anomalous KMT coupling.

Another goal is to include the effect of topological fluctuations on the $U_A(1)$ breaking part of the effective action.  In the meson model framework, the anomaly is described by the KMT determinant.  The corresponding bare coupling, defined at the initial UV scale, $\Lambda$, of the effective action, is determined by the underlying theory of QCD. If $\Lambda$ of the effective theory were high enough, none of the bare couplings could depend on the environment, but in meson models, with $\Lambda$ being ${\cal O}(1 \GeV)$, the temperature dependence of the KMT coupling, generated by QCD dynamics above the scale $\Lambda$, could be important. It is well known that at high temperatures topological fluctuations are well represented by a dilute gas of single charged instantons, which radically cuts down the strength of the anomaly as the temperature rises.  If the temperature decreases, it is assumed that the instanton liquid condenses, which displays only a weak $T$ dependence.  

In this study, we propose a simple interpolating parametrization of the aforementioned temperature dependence, similarly to what has already been employed by several authors \cite{ruivo12,ishii16,ishii17,rai20}. Our main focus will be on investigating how the two competing effects (i.e., mesonic fluctuations and instantons contributions) produce a realistic, environment-dependent anomaly coefficient.

The paper is organized as follows. In Sec. II, based on a chiral invariant expansion, we introduce an approximate effective potential, which is going to be used as an ansatz for the renormalization group calculations. In Sec. III we derive flow equations for the coefficient functions, which replace coupling constants of the ordinary framework. This is to be done in two separate steps, which include different background fields that yield three separate flow equations. These equations need to be solved simultaneously via numerics, which we do in Sec. IV. There we discuss how results change with respect to different assumptions on the bare anomaly coupling and show the thermal behavior of the system (condensate evaporation, mass spectrum, and anomaly evolution). The reader finds the discussion and outlook in Sec. V.

\section{Chiral effective potential}

We are working in an effective theory framework, where the dynamical variables are meson fields,
\bea
M=(s_a+i\pi_a)T^a.
\eea
Here, $s_a$ and $\pi_a$ refer to the scalar and pseudoscalar components, respectively, and $T^a$ are the $U(3)$ generators in the fundamental representation, $\Tr (T^aT^b)=\delta^{ab}/2$. Any effective model built upon the $M$ fields needs to reflect chiral symmetry; i.e., it has to be invariant under the transformations $M \rightarrow L M R^{\dagger}$, where $L$ ($R$) refers to left- (right-)handed $U(3)$ chiral rotations.  

The effective potential of a quantum field theory, $V$, is defined as the zero momentum part of the effective action, $\Gamma$. In this study, we approximate (Euclidean) $\Gamma$ with a standard kinetic term (no wave function renormalization will be taken into account as the anomalous dimension is expected to be small in scalar models \cite{berges02}) plus a local effective potential, V, as
\bea
\label{Eq:Gamma}
\Gamma = \int_ x \Big(\Tr[\partial_i M^\dagger \partial_i M] + V[M]\Big).
\eea
Note that $\Gamma$ and thus $V$ can only contain combinations of the $M$ field that are invariant under $U_L(3)\times U_R(3)$ chiral transformations. For three flavors, there are three independent chiral invariant combinations, that is,
\bea
\label{Eq:inv}
\rho &=& \Tr (M^\dagger M), \nonumber\\
\tau &=& \Tr \big(M^\dagger M - \Tr(M^\dagger M)/3\big)^2, \nonumber\\
\rho_3 &=& \Tr \big(M^\dagger M - \Tr(M^\dagger M)/3\big)^3.
\eea
Obviously, Eq. (\ref{Eq:inv}) is a nonunique set, but once it is chosen, all other chiral invariants can be expressed in terms of $\rho$, $\tau$, and $\rho_3$. When one is doing phenomenology, one also needs combinations that give account of the $U_A(1)$ anomaly. The KMT determinant term, i.e.,
\bea
\label{Eq:invdelta}
\Delta = \det M^\dagger + \det M
\eea
is the prototype of such terms, and it is the only one that is (super)renormalizable in four dimensions. One should be aware that
\bea
\tilde{\Delta} = \det M^\dagger - \det M
\eea
is forbidden due to parity reasons, but $\tilde{\Delta}^2$ could in principle be included. It can be shown, however, that it is not independent,
\bea
\tilde{\Delta}^2 = \Delta^2 - 4\rho^3/27 - 2\rho \tau/3 + 4\rho_3/3.
\eea
That is, if Eqs. (\ref{Eq:inv}) are all included in the potential, then $\Delta$ and its powers are the only operators that describe the $U_A(1)$ anomaly, apart from the obvious fact that any combination of operators (\ref{Eq:inv}) multiplied by $\Delta$ also realizes a $U_A(1)$ breaking operator.

Keep in mind that without explicit symmetry breaking terms, spontaneously broken chiral symmetry shows the pattern $U_L(3)\times U_R(3) \rightarrow U_V(3)$, where the latter index refers to vector transformations (in which the parameters of the left and right ones are equal). That is to say, the ground state is proportional to the unit matrix, $M \sim \hat{{\bf 1}}$,  where both $\tau=0$ and $\rho_3=0$. Note that if we are to treat finite quark masses as perturbations they can only slightly modify this state. Since we are interested in the thermodynamic properties of the system around this vacuum, for the exploration of fluctuating field configurations, it can be assumed that $\rho \neq 0$ but $\tau \approx 0$, $\rho_3 \approx 0$. This leads to the natural choice of expansion in terms of chiral invariants:
\bea
\label{Eq:pot}
V(\rho,\tau,\rho_3,\Delta;H)&=& U(\rho) + C(\rho) \tau + D(\rho) \rho_3 + A(\rho) \Delta \nonumber\\
&-& \Tr \big(H(M+M^\dagger)\big),
\eea
where the last term is the aforementioned explicit symmetry breaking piece, $H=h_0T^0+h_8T^8$ (no isospin breaking is assumed). This expansion is completely analogous to the usual chiral potential of the linear sigma model, but with $\rho$-dependent couplings. Note that, while the $U_V(3)$ vacuum does not suggest an expansion in terms of $\Delta$, we nevertheless do so; see the second-to-last term in the rhs of (\ref{Eq:pot}). The reason for this is that, due to the $\rho$-dependent coefficient, $A(\rho)$, which multiplies $\Delta$ and thus allows the backreaction of the chiral order parameter on the strength of the axial anomaly (\ref{Eq:pot}), already goes way beyond usual perturbative treatments. In such approaches, the $\rho$ dependence of $A(\rho)$ is completely neglected. Nevertheless, it would be very interesting to include higher powers of $\Delta$, as suggested and analyzed in detail in Ref. \cite{pisarski20} for the two-flavor case.

The main goal of this study is to calculate numerically the effective potential (\ref{Eq:pot}) at various $T$ temperatures. This will be done in the FRG framework \cite{wetterich93,morris94}. In the core of the formalism lies the scale-dependent effective action, $\Gamma_k$, and the corresponding local potential, $V_k$. The former differs from $\Gamma$ in the sense that in $\Gamma_k$ infrared fluctuations below momentum $k$ are suppressed. Obviously, $\Gamma_{k=0}=\Gamma$, $V_{k=0}=V$, and we also wish to keep the form of (\ref{Eq:pot}) at all scales and therefore make the coefficient functions $k$ dependent,
\bea
U(\rho) \quad \rightarrow \quad U_k(\rho),  \qquad C(\rho) \quad \rightarrow \quad C_k(\rho), \nonumber\\
D(\rho) \quad \rightarrow \quad D_k(\rho), \qquad A(\rho) \quad \rightarrow \quad A_k(\rho).
\eea
Maintaining the approximate form of (\ref{Eq:Gamma}) for $\Gamma_k$, $V_k$ obeys the finite temperature flow equation
\bea
\label{Eq:flow}
\partial_k V_k = \frac{T}{2} \sum_n \int_{|\vec{q}|<k} \frac{d^3q}{(2\pi)^3} \tilde{\partial}_k \Tr \log (\omega_n^2+ k^2 + V_k''),
\eea
where $V_k''$ is the $18\times 18$ second derivative matrix of $V_k$, $\omega_n=2\pi n T$ are bosonic Matsubara frequencies, and the $\tilde{\partial}_k$ differential operator acts only on the explicit $k$ dependence. Note that (\ref{Eq:flow}) is sometimes called the optimized flow equation \cite{litim01} in the local potential approximation. Also note that, since the integrand is $q$ independent, the momentum integral merely gives a volume factor.

Our first task is to plug the ansatz (\ref{Eq:pot}) of the effective potential into (\ref{Eq:flow}) and extract individual flow equations for the coefficient functions $U_k(\rho)$, $C_k(\rho)$, $D_k(\rho)$, and $A_k(\rho)$. The main problem here is that $V_k''$ in the rhs of (\ref{Eq:flow}) can only be expressed in terms of field variables $s^a$ and $\pi^a$, but by construction, $V_k$ depends on the invariant combinations $\rho$, $\tau$, $\rho_3$, and $\Delta$.  Therefore, it should be emphasized that the explicit expression of the rhs of (\ref{Eq:flow}) should combine into invariants (\ref{Eq:inv}) and (\ref{Eq:invdelta}), consistently with the functional form of $V_k$ appearing in the lhs. It is highly nontrivial how one obtains from the rhs of (\ref{Eq:flow}) a form, which is compatible with (\ref{Eq:pot}).  This is to be done in the next section.

\section{Renormalization group flows}

Solving the flow equation, one needs an initial condition. In the FRG framework, it is assumed that at some $\Lambda$ UV scale the potential is known, and one integrates the flow equation down toward $k\rightarrow 0$. The usual assumption is that $V_{k=\Lambda}$ takes the form of a classical potential, i.e.,  it includes only operators that are allowed by (perturbative) renormalizability, with coefficients that are environment independent being some functions of $\Lambda$. As already announced in the Introduction, if we are dealing with effective models, and $\Lambda$ is not large enough, this assumption may be altered due to the environment dependence of the interactions at higher scales. We will come back to this point in the next section. In accordance with (perturbative) renormalizability, pieces of the effective potential in the UV are assumed to be the following:
\bea
\label{Eq:incon}
U_\Lambda(\rho) &=& m^2\rho + g_1 \rho^2, \quad C_\Lambda (\rho) = g_2,  \nonumber\\
D_\Lambda(\rho) &=& 0, \quad A_\Lambda(\rho) = a.
\eea
The parameters $m^2$, $g_1$, $g_2$ and $a$ can be determined using physical input, e.g. the mass spectrum calculated from the effective potential at $k=0$.  Note that the RG flow equations are fully determined by the dimensionality and the symmetry of the system; thus, $H$ does not enter to the rhs of the flow equation. That is equivalent to saying that none of the flows is sensitive to the explicit breaking and they cannot generate terms that break chiral symmetry. That is, $H$ remains a $k$-independent constant matrix at any scale.

In what follows, we show how to extract flow equations for the coefficient functions. Since $\rho_3$ is a nonrenormalizable operator, it is expected that its effect is small; therefore, in our analysis, we set $D_k\equiv 0$ for all $k$. Note that throughout the calculations one has to be consistent with this assumption; i.e., no $\rho_3$ dependence should be generated in the RG flow. The lhs of the flow equation (\ref{Eq:flow}), therefore, does not contain $\rho_3$, and it becomes
\bea
\partial_k U_k(\rho) + \partial_k C_k(\rho) \tau + \partial_k A_k(\rho) \Delta.
\eea
Now, as already mentioned in the previous section, the problem with extracting expressions for $\partial_k U_k$, $\partial_k C_k$, and $\partial_k A_k$ is that $V_k''$ in the rhs of (\ref{Eq:flow}) can be expressed in terms of the fields and not the invariants. Obviously, the flow equation is chirally symmetric; therefore, these field dependences must eventually be combined into invariant tensors, but from a practical point of view, it is highly nontrivial how to perform the calculations. One has to invent an expansion in terms of the field variables that generate an expression in the rhs of (\ref{Eq:flow}) that is compatible with (\ref{Eq:pot}).

For this, one can exploit the obvious feature that the expression of both sides of (\ref{Eq:flow}) in terms of the invariants is unique. That is, they are reconstructed from a multitude of specific field configurations when evaluating the rhs of (\ref{Eq:flow}), which share the feature that the $\rho,\tau,\Delta$ invariants can be disentangled in a unique fashion. One may choose the most convenient background, making the reconstruction of the invariants the simplest. Once all flow equations are set up, one analyzes the emerging potential in the actual ``physical'' background (in our case $M=s_0T^0+s_8T^8$), dictated by the direction of the linear explicit breaking [last term in (\ref{Eq:pot})].

\subsection{Flows of $U_k$ and $A_k$}

In this subsection, we work with the background that is defined by $M=(s_0+i\pi_0)T^0$. The main advantage of this choice is that in such configurations $\tau\equiv 0$ (and also the omitted $\rho_3=0$), while the remaining invariant combinations are given by
\bea
\label{Eq:invpi}
\rho=\frac12(s_0^2+\pi_0^2), \quad \Delta = \frac{s_0^3-3\pi_0^2s_0}{3\sqrt6}.
\eea
That is to say, no dependence on $\tau$ appears in the lhs of (\ref{Eq:flow}) and thus one is able to extract the flows of $U_k$ and $A_k$ as the pure $\rho$-dependent, and the ${\cal O}(\Delta)$ parts of the rhs of (\ref{Eq:flow}), respectively. This is rather convenient, since one can perform calculations without the need of keeping track of the identification of the $\tau$ invariant. Note that, as a result, the flow of $C_k(\rho)$ cannot be obtained in this background, but as we will see, it does contribute to both the flows of coefficients $U_k(\rho)$ and $A_k(\rho)$.

Since the background we are working with is proportional to the unit matrix, symmetry of the fluctuations around this configuration requires the mass matrix, $V_k''$, to have eight degenerate doublet eigenmodes, corresponding to the planes $\{s_i,\pi_i\}$, $i=1,2,...,8$, and one different doublet eigenmode in the $\{s_0,\pi_0\}$ plane (see Appendix A for the calculation of $V_k''$).  Using the identity ``$\Tr \log = \log \det$'' and the notation $\Omega=\int_{|\vec{q}|<k}\frac{d^3 q}{(2\pi)^3}=\frac{k^3}{6\pi^2}$ for the volume factor, the rhs of (\ref{Eq:flow}) yields 
\begin{widetext}
\bea
\label{Eq:flowrhs1}
8&\times& \frac{\Omega}{2} T \sum_n \tilde{\partial}_k \log \Big((\omega_n^2+k^2+U_k'+A_k'\Delta)^2 +\frac43(\omega_n^2+k^2+U_k'+A_k'\Delta)\rho C_k-\frac13\rho A_k^2+2A_kC_k\Delta\Big)\nonumber\\
&+&\frac{\Omega}{2}T\sum_n \tilde{\partial}_k \log \Big((\omega_n^2+k^2+U_k'+A_k'\Delta)^2 +2(\omega_n^2+k^2+U_k'+A_k'\Delta)\big(3A_k'\Delta+(U_k''+A_k''\Delta)\rho\big)\nonumber\\
&&\hspace{2.4cm}-6A_k\Delta(U_k''+A_k''\Delta)-\frac43\rho(A_k+\rho A_k')^2+9\Delta^2A_k'^2\Big),
\eea
\end{widetext}
where the first term comes from the $s-\pi$ mixing in the $i=1,2,...8$ sectors, while the second one is obtained from the doublet of $i=0$. Note that the $\rho$ and $\Delta$ invariants could already be identified within each sector individually. To transform (\ref{Eq:flowrhs1}) into a compatible form with the ansatz of (\ref{Eq:pot}), one has to expand (\ref{Eq:flowrhs1}) to linear order in $\Delta$. This yields
\bea
\label{Eq:flowU}
&&\partial_k U_k(\rho) = \frac{\Omega}{2}T\sum_n \tilde{\partial}_k (8\log D_8 + \log D_0)
\eea
for the flow equation of $U_k$ and
\bea
\label{Eq:flowA}
\partial_k A_k(\rho) &=&\hspace{0.05cm} \Omega T\sum_n \tilde{\partial}_k \Big[\frac{8}{D_8}\Big(A_k'(\omega_n^2+k^2+U_k')\nonumber\\
&&\hspace{2.4cm}+\frac23\rho C_kA_k'+A_kC_k\Big)\nonumber\\
&&\hspace{1cm}+\frac{1}{D_0} \Big((4 A_k'+\rho A_k'')(\omega_n^2+k^2+U_k') \nonumber\\
&&\hspace{1cm}+U_k''(\rho A_k'-3A_k)\Big)\Big] \nonumber\\
\eea
for that of $A_k$. Here,
\begin{subequations}
\bea
\label{Eq:D8}
D_8&=&(\omega_n^2+k^2+U_k')(\omega_n^2+k^2+U_k'+\frac43 \rho C_k)\nonumber\\
&-&\frac13 \rho A_k^2, \\
\label{Eq:D0}
D_0&=&(\omega_n^2+k^2+U_k')(\omega_n^2+k^2+U_k'+2\rho U_k'') \nonumber\\
&-&\frac43\rho(A_k+\rho A_k')^2.
\eea
\end{subequations}
Expanding the rhs of (\ref{Eq:flowU}) and (\ref{Eq:flowA}) in terms of the anomaly function, at the next-to-leading order, one recovers the results of Ref. \cite{fejos16}. Note that (\ref{Eq:flowU}) can also be obtained directly by choosing the imaginary background $M=i\pi_0T^0$, in which in addition to $\tau = 0$ also $\Delta = 0$.  Calculating the $V_k''$ matrix elements in such background and plugging it into (\ref{Eq:flow}), one arrives directly at (\ref{Eq:flowU}).

\subsection{Flow of $C_k$}

For the determination of the flow equation of $C_k$, the $M=i(\pi_0T^0+\pi_8T^8)$ purely imaginary background appears to be the most convenient. In this case, the cubic invariant $\Delta$ automatically vanishes, and we have
\bea
\rho = \frac12(\pi_0^2+\pi_8^2), \quad \tau=\frac13 \pi_8^2\Big(\pi_0-\frac{1}{2\sqrt2}\pi_8\Big)^2.
\eea
In the applied background, the fluctuation determinant breaks into three degenerate doublets in the $\{\s_i,\pi_i\}$, $i=1,2,3$ planes, four degenerate doublets in the $\{\s_i,\pi_i\}$, $i=4,5,6,7$ planes, and a fully coupled quartet in the subspace $\{s_0,s_8,\pi_0,\pi_8\}$. The former seven $2\times 2$ subsectors can be calculated quite easily, but the complete analytic evaluation of the $4\times 4$ determinant is a lot more messy. For the rhs of (\ref{Eq:flow}), one arrives at
\begin{widetext}
\bea
\label{Eq:rhsCk}
&&3\times \frac{\Omega}{2} T\sum_n \tilde{\partial}_k \log \bigg[(\omega_n^2+k^2+U_k'+C_k'\tau)^2+(\omega_n^2+k^2+U_k'+C_k'\tau)\Big(\frac43 \rho+4\sqrt{\frac23}\sqrt{\tau}\Big)C_k\nonumber\\
&&\hspace{3cm}+2C_k^2\Big(\tau+\frac23\sqrt{\frac23}\rho\sqrt{\tau}\Big)-A_k^2\Big(\frac13\rho-\sqrt{\frac23}\sqrt{\tau}\Big)\bigg]\nonumber\\
&+&4\times \frac{\Omega}{2}T\sum_n \tilde{\partial}_k \log \bigg[(\omega_n^2+k^2+U_k'+C_k'\tau)^2+(\omega_n^2+k^2+U_k'+C_k'\tau)\Big(\frac43\rho-2\sqrt{\frac23}\sqrt{\tau}\Big)C_k\nonumber\\
&&\hspace{3cm}+2C_k^2\Big(\tau-\frac13\sqrt{\frac23}\rho\sqrt{\tau}\Big)-A_k^2\Big(\frac13\rho+\sqrt{\frac16}\sqrt{\tau}\Big)\bigg]\nonumber\\
&+&\frac{\Omega}{2}T\sum_n \tilde{\partial}_k \log \Big[D_0 D_8-4\sqrt{\frac23}\Big(\frac{A_k^2}{4}+(\omega_n^2+k^2+U_k'+C_k \rho/3)C_k\Big)D_0\sqrt{\tau}+F\pi_8^2)\Big],
\eea
\end{widetext}
where $F$ is a complicated function of $\pi_0$ and $\pi_8$. Note that as opposed to the cases of $U_k$ and $A_k$, now in none of the determinants do the field variables combine into invariants, as reflected by the merely formal appearance of the nonanalytic terms $\sim \sqrt{\tau}$. These need to be canceled out, and eventually indeed do so. To see that one expands (\ref{Eq:rhsCk}) in terms of $\pi_8$, which at ${\cal O}(\pi_8^0)$ reproduces the flow equation for $U_k$,  at ${\cal O}(\pi_8)$ shows that all contributions exactly cancel (which is equivalent of saying that all the formal $\sim \sqrt{\tau}$ terms drop), while at ${\cal O}(\pi_8^2)$ one evaluates $F$ at $\pi_8=0$ and gets $F = F_0 \pi_0^2 + {\cal O}(\pi_8)$ with $F_0$ still $\pi_0$ (and thus $\rho$) dependent. Then, the identification of the $\tau = \pi_0^2\pi_8^2/3 + {\cal O}(\pi_8^3)$ invariant is straightforward, and it leads to the flow equation for $C_k$
\begin{widetext}
\bea
\partial_k C_k&=&\Omega T\sum_n\tilde{\partial}_k\Bigg\{\frac{7}{2D_8}\left(2C_k'(\omega_n^2+k^2+U_k^\prime)+\frac{4}{3}\rho C_kC_k'+2C_k^2\right)+\frac{2}{D_8}\left(\frac{3}{2}C_k'(\omega_n^2+k^2+U_k^\prime)+\frac{1}{3}\rho C_kC_k'-\frac{1}{4}A_kA_k'\right)
 \nonumber\\
&-&\frac{2}{3D_8^2}\Big(A_k^2+\frac{4}{3}\rho C_k^2+4C_k(\omega_n^2+k^2+U_k')\Big)^2+\frac{1}{D_0}\Big((3C_k'+\rho C_k'')(\omega_n^2+k^2+U_k')+\frac{3}{2}A_k'(A_k+\rho A_k')+\rho C_k' U_k''\Big) \nonumber\\
&-&\frac{4}{3D_8^2}\Bigg(\frac{1}{16}A_k^4+\frac{7}{12}\rho A_k^2C_k^2+\frac{2}{9}\rho^2C_k^4 (\omega_n^2+k^2+U_k')\Big(A_k^2+\frac{1}{3}\rho C_k^2\Big)C_k+\frac{5}{4}(\omega_n^2+k^2+U_k')^2C_k^2\Bigg) \nonumber
\eea
\bea
\label{Eq:flowC}
&-&\frac{8}{D_0D_8}\Bigg((\omega_n^2+k^2+U_k')^2\Big(\frac{5}{12}C_k^2+\frac{3}{16}(U_k''+\frac{4}{3}\rho C_k')^2+\frac{1}{2}C_k(U_k''+\frac{4}{3}\rho C_k')\Big)\nonumber\\
&&\hspace{1.1cm}+(\omega_n^2+k^2+U_k')\Big(\frac{1}{6}\rho C_k^2(U_k''+\frac{2}{3}C_k')+\frac{1}{16}(U_k''+\frac{4}{3}\rho C_k')(A_k^2-4\rho A_kA_k'-4\rho^2A_k'^2)+\frac{C_k}{24}(3A_k^2-4\rho^2A_k'^2)\Big)\nonumber\\
&&
\hspace{1.1cm}+\frac{2}{9}\rho^2U_k''C_k^3-\frac{1}{9}\rho C_k^2(A_k^2-\rho A_kA_k'-2\rho^2 A_k'^2)-\frac{1}{4}\rho C_kA_k^2U_k''\nonumber\\
&&\hspace{1.1cm}-\frac{2}{9}\rho^2 C_kC_k' A_k(A_k+\rho A_k')-\frac{A_k}{48}(A_k+\rho A_k')(A_k^2-4\rho^2 A_k'^2)\Bigg)\nonumber\\
&+&\frac{A_k^2}{6D_0D_8}\Big(4C_k(\omega_n^2+k^2+U_k')-A_k^2\Big)+\frac{(\omega_n^2+k^2+U_k')^2A_k^2}{4D_0D_8^2}\Big(A_k^2+\frac{8}{3}\rho A_kA_k'+\frac{4}{3}\rho^2A_k'^2-\frac{8}{3}\rho C_k(U_k''-\frac{2}{3}C_k)\Big)\nonumber\\
&+&\frac{(\omega_n^2+k^2+U_k')}{4D_0D_8}\Bigg(6(\omega_n^2+k^2+U_k')(U_k''-\frac{2}{3}C_k)^2-\frac{2A_k^2}{D_8}(\omega_n^2+k^2+U_k')^2(U_k''-\frac{2}{3}C_k)\nonumber\\
&&\hspace{2.5cm}+(A_k^2+\frac{8}{3}\rho A_kA_k'+\frac{4}{3}\rho^2A_k'^2)\Big(\frac{4\rho C_kA_k^2}{3D_8}-3(U_k''-\frac{2}{3}C_k)\Big)\Bigg)\nonumber\\
&+&\frac{1}{D_0}\Bigg(A_kA_k'+\rho A_k'^2\left(\frac{1}{2}-\frac{(\omega_n^2+k^2+U_k')^2}{D_8}\right)\nonumber\\
&&\hspace{0.8cm}-\frac{\omega_n^2+k^2+U_k'}{4D_8}\Big(
A_k^2(2C_k+U_k'')-4\rho A_kA_k'(U_k''-\frac{2}{3}C_k)+4\rho^2A_k'^2(U_k''+\frac{2}{3}C_k)\Big)\Bigg)\Biggr\},
\eea
\end{widetext}
where the background-independent definitions of $D_8$ and $D_0$ can be read off from (\ref{Eq:D8}) and (\ref{Eq:D0}), respectively. The technical difficulty of the calculation can be illustrated by realizing that the first two terms of (\ref{Eq:rhsCk}) only give the first three contributions in (\ref{Eq:flowC}), the remaining ones come directly from the $F_0$ factor. Finally, we note that, since in the applied background $\rho_3 = {\cal O}(\pi_8^3)$, the outlined calculations do not get contaminated by the appearance of the $\rho_3$ invariant.
 
\section{Numerical results}

Now we solve the coupled differential equations ({\ref{Eq:flowU}), (\ref{Eq:flowA}), and (\ref{Eq:flowC}) using the grid method. We set up three grids in $\rho$ space with spacing $\delta \rho= 50\MeV^2$. All $\rho$ derivatives are calculated using the six-point formula, except close to the grid boundaries, where the five- and four-point formulas have been used. The differential equations are then integrated using the fourth-order Runge-Kutta method, starting from $k=\Lambda\equiv 1 \GeV$ toward $k=0$, using (\ref{Eq:incon}) as initial conditions. As reported in several papers in the literature, the flows slow down when approaching $k\rightarrow 0$, needing gradually more computational time to perform the next step in $k$. We therefore stop them at $k_{\en}= 10\MeV$, at which all functions are practically converged and none of the results is $k$ dependent. The Matsubara sums are performed analytically in (\ref{Eq:flowU}) and (\ref{Eq:flowA}) (see also Appendix B) and numerically in (\ref{Eq:flowC}). In the latter, cutoffs in the sums are chosen such that the final results practically do not depend on their actual value. This meant typically summing up ${\cal O}(1000)$ terms.

The first task before obtaining any result is the parametrization of the model. The $H$ matrix, i.e. its $h_0$ and $h_8$ components, are determined by the partially conserved axial-vector current relations. They read
\bea
m_\pi^2 f_\pi &=& \sqrt{\frac23}h_0+\sqrt{\frac13}h_8, \nonumber\\
m_K^2f_K &=& \sqrt{\frac23}h_0-\frac12\sqrt{\frac13}h_8,
\eea
which gives
\bea
h_0&=&\frac{1}{\sqrt{6}}m_{\pi}^2f_{\pi}+\sqrt{\frac23}m_K^2f_K, \nonumber\\
h_8&=&\frac{2}{\sqrt3}m_{\pi}^2f_{\pi}-\frac{2}{\sqrt3}m_K^2f_K,
\eea
where $m_\pi$ and $m_K$ are the pion and kaon masses, respectively, while $f_\pi$ ($93 \MeV$) and $f_K$ ($113 \MeV$) are the corresponding decay constants.Because of this choice of the $H$ symmetry breaking matrix, in the vacuum, we have a two-component condensate: $M=s_0T^0+s_8T^8$. The remaining four parameters, i.e., $m^2, g_1, g_2, a$, are chosen such that the pseudoscalar masses ($\pi$, $K$, $\eta$, and $\eta'$) reproduce the physical spectrum, i.e., $m_\pi \approx 140 \MeV$, $m_K\approx 494 \MeV$, $m_\eta \approx 548 \MeV$, and $m_{\eta'} \approx 958 \MeV$. The applied parameter set can be seen in Table I. The scalars are expected to be associated with the $f_0(500)$ ($\sigma$), $K_0^*(800)$ ($\kappa$), $a_0(980)$,  and $f_0(980)$ mesons. The masses of the latter excitations turn out to be less accurate, especially that of the $\sigma$ meson. Note that the $\sigma$ field itself is the order parameter of the chiral symmetry breaking, which, in turn, does show the expected vacuum and thermal behavior, despite the unsatisfactory account of its fluctuations. One of the reasons of the former inaccuracies could be that in the Euclidean framework of the RG flow equations the lifetime of the mesons cannot be extracted and treating these fairly broad resonances as stable excitations is a crude approximation. We could have sacrificed some parts of the pseudoscalar spectrum to gain more accurate values for the scalars, but since we are mainly focusing on the anomaly evolution in this study, on top of the lightest pseudoscalars ($\pi$ and $K$), we wished the $\eta$-$\eta'$ system to be as accurate as possible. Also, we note that, even with a more accurate scalar sector (and thus a less accurate pseudoscalar one), the tendency observed for the anomaly evolution would have remained qualitatively the same. 

\begin{table}[t]
\centering
\vspace{0.2cm}
  \begin{tabular}{ c | c }
    $m^2$ & $-0.9 \GeV^2$  \\ \hline
    $g_1$ & 20 \\ \hline
    $g_2$ & 360 \\ \hline
    $a$ & $-2.6 \GeV$ \\ \hline  
    $h_0$ & $ (285 \MeV)^3$ \\ \hline  
     $h_8$ & $(-310 \MeV)^3$ \\
  \end{tabular}
  \caption{Parameters for the initial potential in the UV.}
\end{table}

\begin{figure}[t]
\includegraphics[bb = 0 150 495 570,scale=0.36,angle=-90]{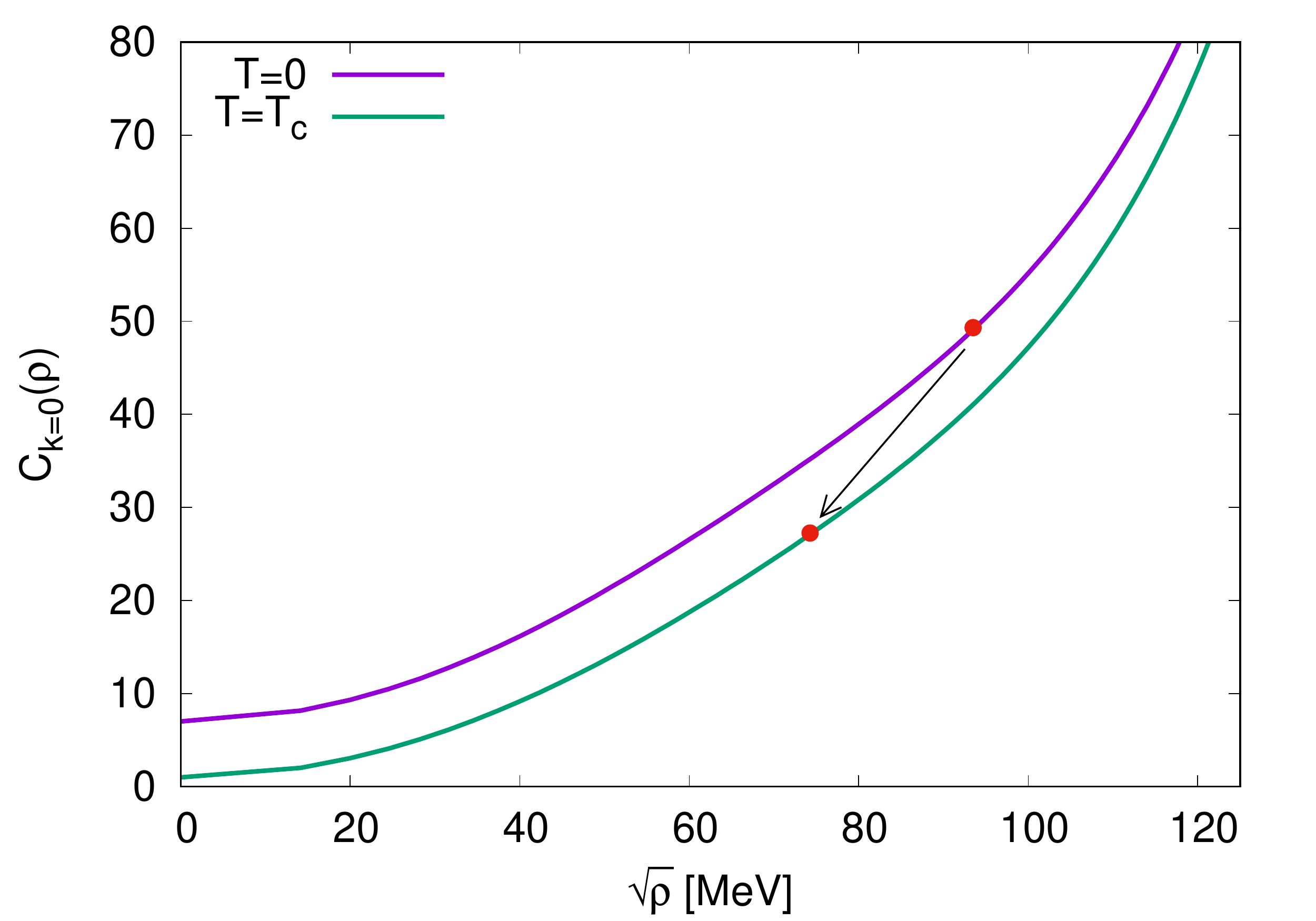}
\caption{Structure of the $C(\rho)$ coefficient function at $T=0$ and at $T=T_c$. Red dots show the value of $C$ corresponding to the actual minimum of the complete effective potential.}
\label{Fig:C}
\end{figure}  

\begin{figure}[t]
\includegraphics[bb = 0 150 495 570,scale=0.36,angle=-90]{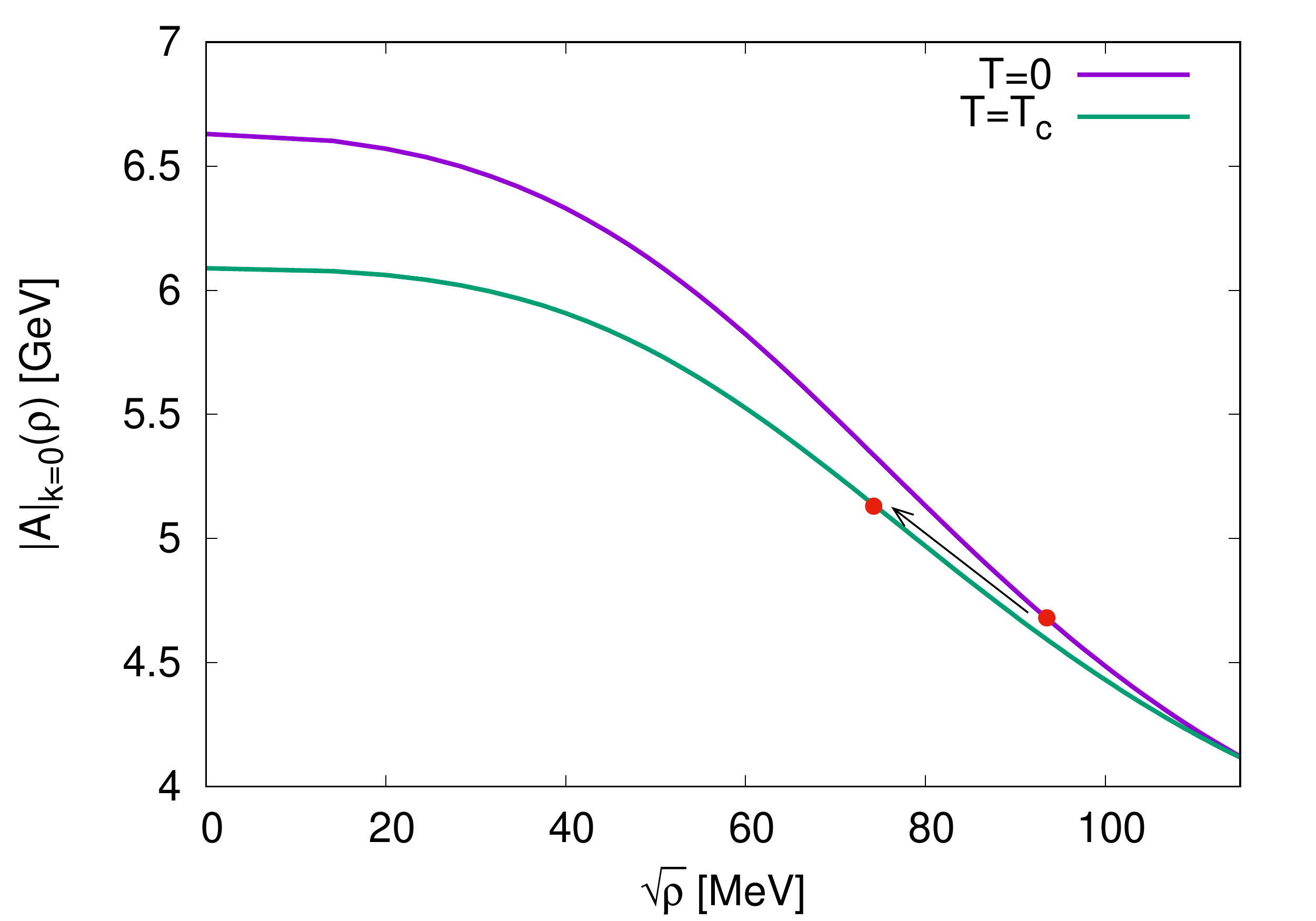}
\caption{Structure of the $A(\rho)$ coefficient function at $T=0$ and at $T=T_c$. Red dots show the value of $|A|$ corresponding to the actual minimum of the complete effective potential.}
\label{Fig:A}
\end{figure}  

\subsection{Zero temperature results}

First, let us review our results at zero temperature. The solution for the $U(\rho)$ function behaves similarly as the effective potential in $O(N)$-like theories; i.e. the symmetry breaking potential gradually flattens as $k$ decreases so that $U(\rho)$ obeys convexity in the IR. The $C(\rho)$ and $A(\rho)$ functions are more interesting. As for the former, in Fig. \ref{Fig:C}, we see that its bare, field-independent value at the UV scale ($g_2=360$) substantially gets modified approaching the IR. This is of no surprise as (based on dimensional analysis) $\sim \log \Lambda^2$ terms supposedly alter the bare coupling, but it is most important to realize that the field dependence of $C(\rho)$ is non-negligible. This shows that resummation in $\rho$, realized by the FRG method, is rather important because, for instance, perturbation theory would definitely not be able to reproduce such behavior. As for $A(\rho)$, we observe a similar pattern, and in accordance with Ref. \cite{fejos16}, its absolute value turns out to be a monotically decreasing function.  Comparing Fig. \ref{Fig:A} with the earlier results of Ref. \cite{fejos16},  now a more moderate tendency is observed in the same sense. 

At $T=0$, for the $[|A|(\rho=0)-|A|(\rho=\rho_{\min})]/|A|(\rho=0)$ ratio now we get around $\sim 30\%$, as opposed to the earlier attempt \cite{fejos16}, where the same quantity was roughly $\sim 40\%$. The rather crude approximation of Ref. \cite{fejos16} somewhat overestimates the field dependence of $A(\rho)$.

The decreasing nature of $|A|(\rho)$ already suggests that, once thermal fluctuations are taken into account, the anomaly strength will increase, since if the chiral condensate evaporates the actual value of $|A|(\rho=\rho_{\min})$, corresponding to the minimum point of the effective potential, becomes larger; see the illustration also in Fig. \ref{Fig:A}. We emphasize the nonperturbative nature of this backreaction of the condensate on the KMT coupling.

\subsection{Finite temperature results}

Minimizing the effective potential with respect to $s_0$ and $s_8$ gives the thermal evolution of the condensates. Instead of the $s_0, s_8$ variables, we use the nonstrange-strange basis:
\bea
\label{Eq:basis}
\begin{pmatrix}
s_{\ns} \\ s_{\s}
\end{pmatrix}
= \frac{1}{\sqrt3} \begin{pmatrix}
\sqrt2 & 1 \\
1 & -\sqrt2
\end{pmatrix}
\begin{pmatrix}
s_0 \\ s_8
\end{pmatrix},
\eea
and denote the minimum points of $V$ as $v_{\ns}, v_{\s}$, thus $\rho_{\min}=(v_{\ns}^2+v_{\s}^2)/2$. Results are shown in Fig. \ref{Fig:cond}. The pseudocritical temperature, $T_c$, is defined through the inflection of the $v_{\ns}(T)$ curve. It comes out surprisingly close to lattice results, we obtain $T_c \approx 158 \MeV$. Here, we see a huge improvement compared to Ref. \cite{fejos16}, in which $T_c$ was off by about a factor of 2. Notice that the strange component evaporates much slower, and its inflection point can be found at a slightly lower (by $\sim 10 \MeV$) value. We also note that one could also use the temperature dependence of the $m_{\sigma}-m_{\pi}$ mass difference to extract the pseudocritical temperature of the transition of the nonstrange condensate. The corresponding inflection point is found at $T_c \approx 167 \MeV$, which is about $5\%$ higher than that obtained from thermal evolution of the nonstrange condensate. This modest variation in the characteristic transition temperature values is fairly compatible with the physics of a smooth crossover.
\begin{figure}[t]
\includegraphics[bb = 0 150 495 570,scale=0.36,angle=-90]{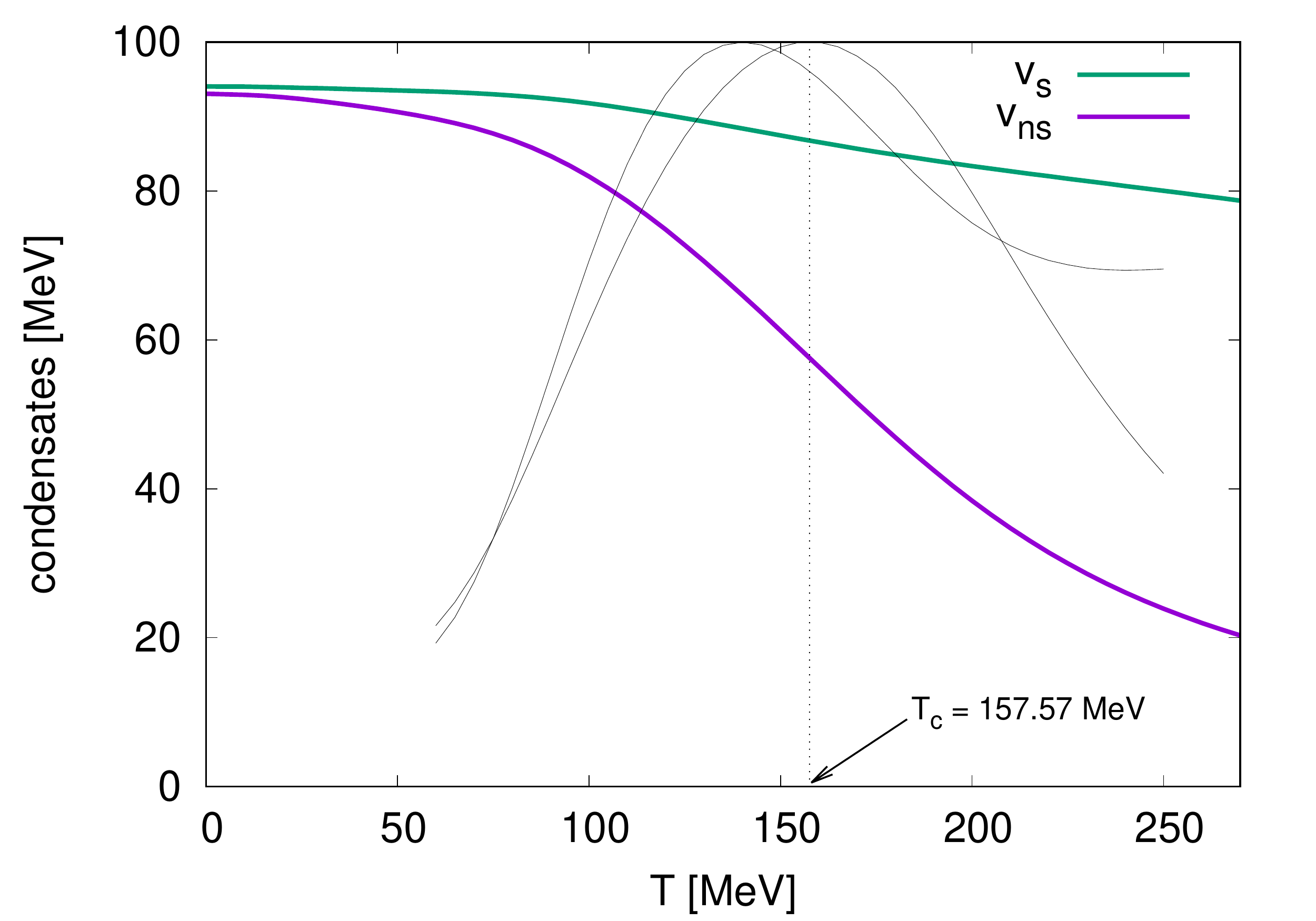}
\caption{The nonstrange and strange condensates as a function of the temperature ($T$). The black curves show the $T$-derivative of each condensate, normalized to the top of the figure. Our result for the pseudocritical temperature is $T_c \approx 158 \MeV$.}
\label{Fig:cond}
\end{figure} 
\begin{figure}[t]
\includegraphics[bb = 0 150 495 570,scale=0.36,angle=-90]{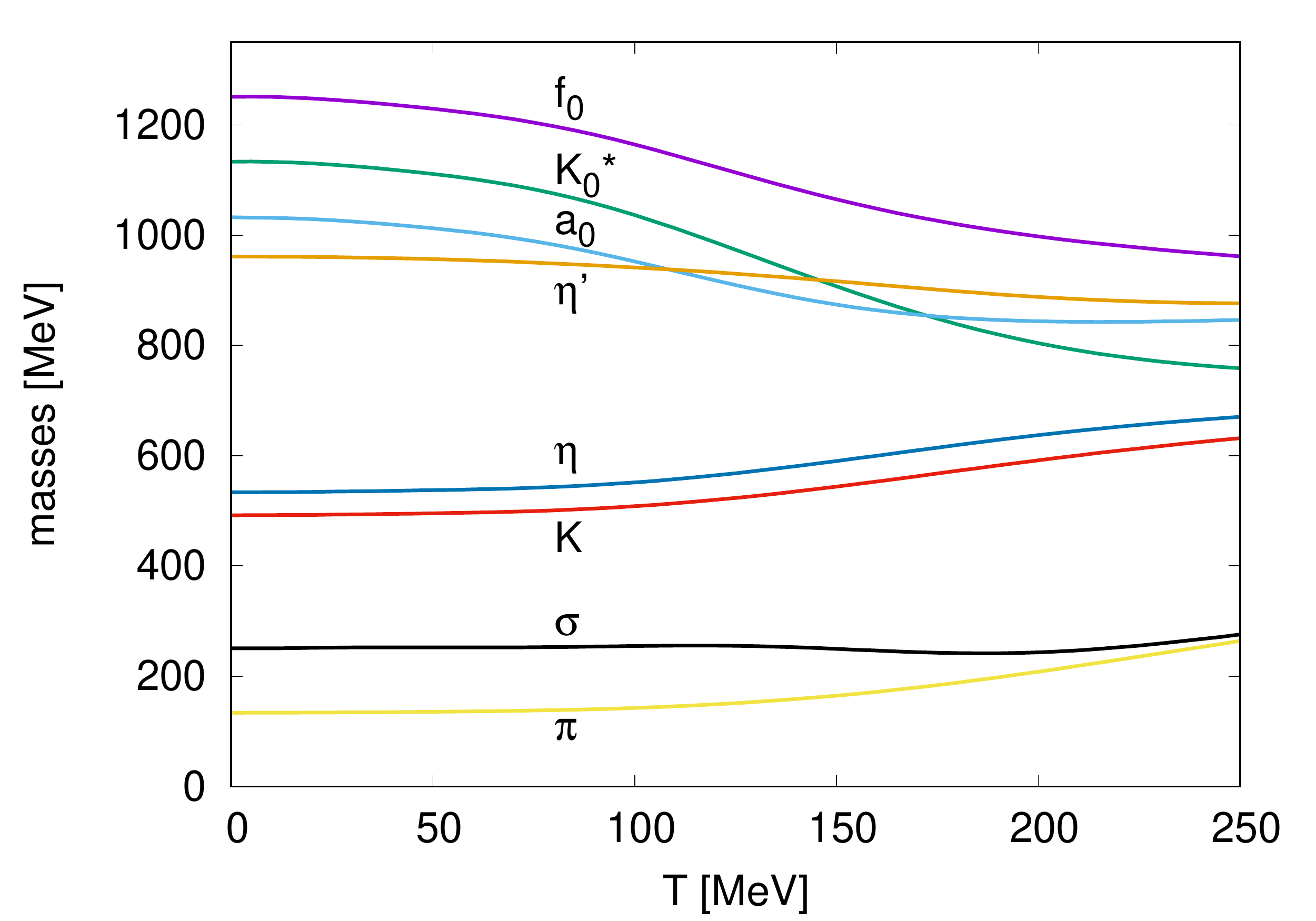}
\caption{The mass spectrum at finite temperature, where the bare anomaly parameter is temperature independent.}
\label{Fig:mass}
\end{figure}  

In Fig. \ref{Fig:mass}, we show the thermal evolution of the mass spectrum, calculated from the second derivative of the effective potential; see the details in Appendix A. As announced at the beginning of this section, for parametrization, the pseudoscalar masses were used. In such parametrization the scalar spectrum seems to be less accurate; in particular, a rather small $\sigma$ and a high $f_0$ mass can be found. It is an open question whether a more sophisticated treatment of the RG flows may cure the scalar spectrum, and it will be investigated in a separate study. We draw attention to the $\eta'$ mass as the temperature rises, which shows no drop toward the pseudocritical temperature. This already hints that the $U_A(1)$ anomaly does not seem to get restored. Note that one needs to be careful with drawing conclusions on the anomaly behavior solely from the mass spectrum, as it contributes typically to masses through ``anomaly strength $\times$ condensate'' type terms, which can also drop solely from condensate evaporation, while the $U_A(1)$ symmetry is still being broken. The temperature dependence of the anomaly coefficient, $A(\rho=\rho_{\min})$, defined at the minimum point of $V$ is displayed explicitly in Fig. \ref{Fig:anom}. The figure shows what has already been expected from Fig. \ref{Fig:A}; through mesonic fluctuation effects, the anomaly gets larger with respect to the temperature.

\subsection{Instanton contributions}

Obviously, we are not at the end of the story. Beyond $T_c$, the $U_A(1)$ symmetry has to be restored, as shown by the semiclassical approximation of the instanton tunelling amplitude. In the instanton liquid model, the topological susceptibility is approximated via the instanton density, $\chi_{\top} \simeq n(r)$, where $r$ is the average instanton size \cite{schaefer96,schaefer98}. At temperatures significantly higher than $T_c$, $n(r)$ contains an exponential suppression factor,
\bea
\label{Eq:nr}
n(r) \sim \exp [-8(\pi r T)^2/3].
\eea
Keeping in mind the effective meson model we are working with, if the bare KMT coupling, $a$, is proportional to $\chi_{\top}$, then one is able to reproduce the Witten-Veneziano relation \cite{schaefer96,schaefer98}. As already pointed out in Ref. \cite{schaefer98}, however, it is not entirely correct to associate the topological susceptibility with either the KMT coupling, or with the instanton density. We still find it phenomenologically the most reasonable to use an interpolating form, following the $T=0$ and the asymptotically large-$T$ behavior of the topological fluctuations \cite{ruivo12,ishii16,ishii17,rai20}. Therefore, on top of the already discussed scenario, where the bare anomaly coupling, $a$, is temperature independent, we explore three different assumptions, for which $a(T)$ does depend on the temperature \cite{ruivo12,rai20,ishii16,ishii17}:
\bea
i)\quad a(T) &=& a_0 \exp [-8(\pi r T)^2/3], \nonumber\\
ii)\quad  a(T) &=&
\begin{cases}
a_0, \hspace{3.5cm} \ife \hspace{0.1cm} T<T_c\\
a_0 \exp [-8(\pi r)^2 (T^2-T_c^2)/3] , \hspace{0.2cm} \els
\end{cases}\nonumber\\
iii)\quad  a(T) &=&
\begin{cases}
a_0, \hspace{3.5cm} \ife \hspace{0.1cm} T<T_0\\
a_0 \exp [-8(\pi r)^2 (T^2-T_0^2)/3] , \hspace{0.2cm} \els.
\end{cases}\nonumber
\eea
Assumption $i)$ is rather crude, as the exponential suppression should take place at very high $T$, well beyond $T_c$. Case $ii)$ is more reasonable, as it is sometimes argued that it is the Debye screening effect of the instanton field that causes the exponential suppression and, therefore, it does not affect the instanton density below $T_c$. However, as it turns out, such an approximate $a(T)$ function makes the actual critical temperature grow; therefore, it seems more appropriate to introduce a $T_0$ parameter, which should be tuned such that $T_c$ retains its physical value. This defines scenario $iii)$.

\begin{figure}[t]
\includegraphics[bb = 0 150 495 570,scale=0.36,angle=-90]{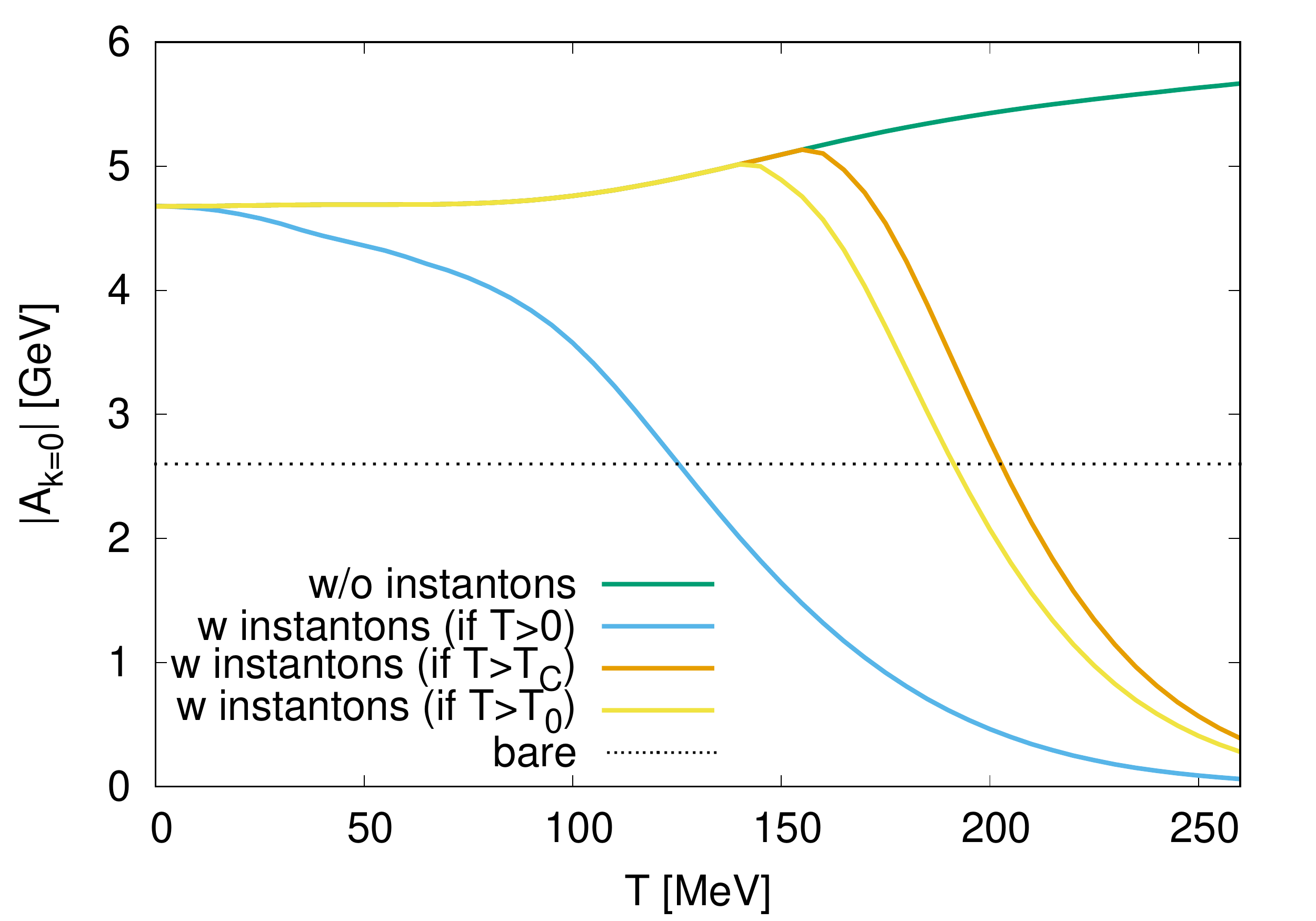}
\caption{Behavior of the dressed anomaly parameter, evaluated at the minimum point of the effective potential, as a function of the temperature. For explanation of the various scenarios see the text.}
\label{Fig:anom}
\end{figure}  

\begin{figure}[t]
\includegraphics[bb = 0 150 495 570,scale=0.36,angle=-90]{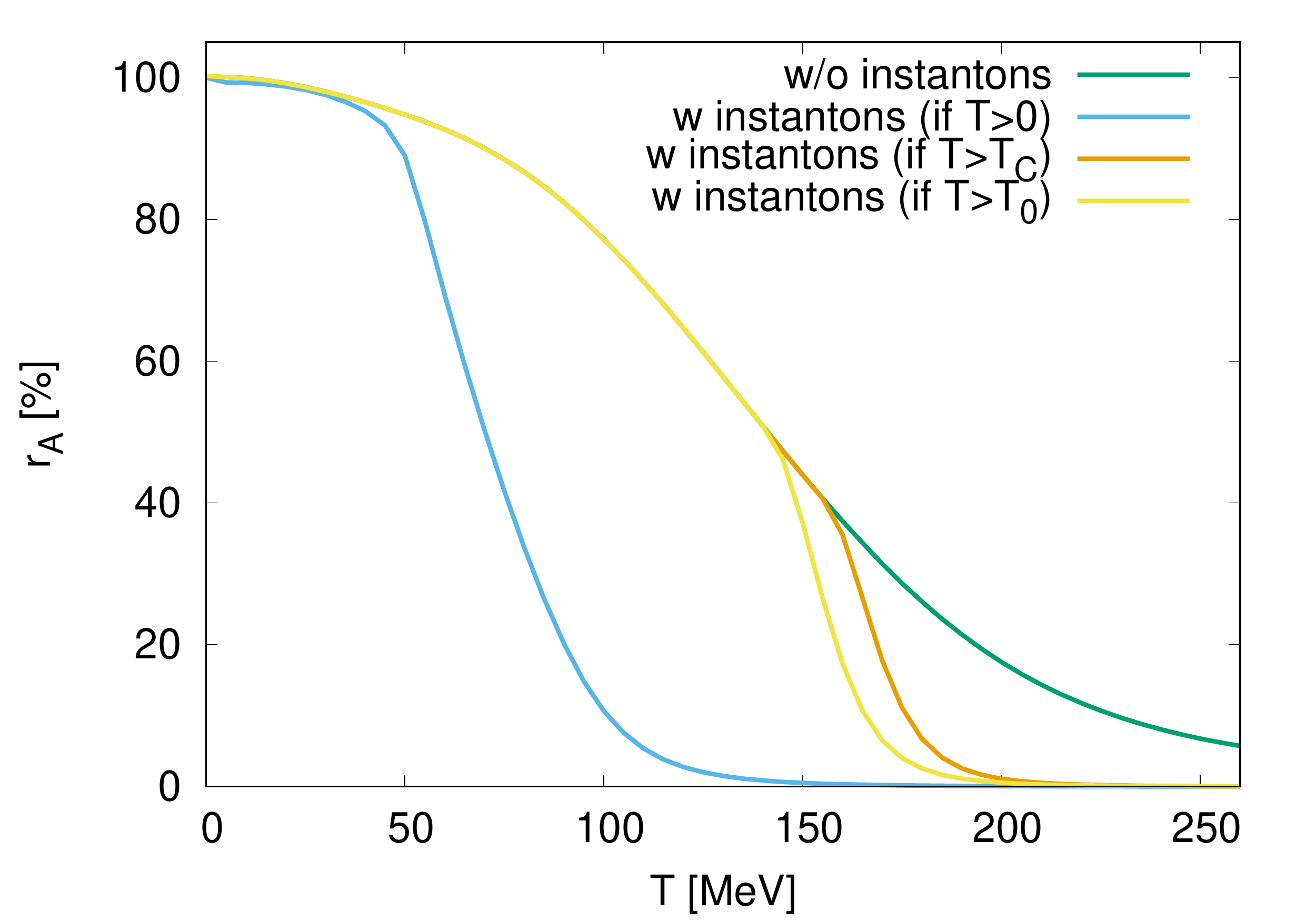}
\caption{The $r_A$ ratio as a function of the temperature for various scenarios for the bare anomaly parameters.}
\label{Fig:anom2}
\end{figure}

\begin{figure}[t]
\includegraphics[bb = 0 150 495 570,scale=0.36,angle=-90]{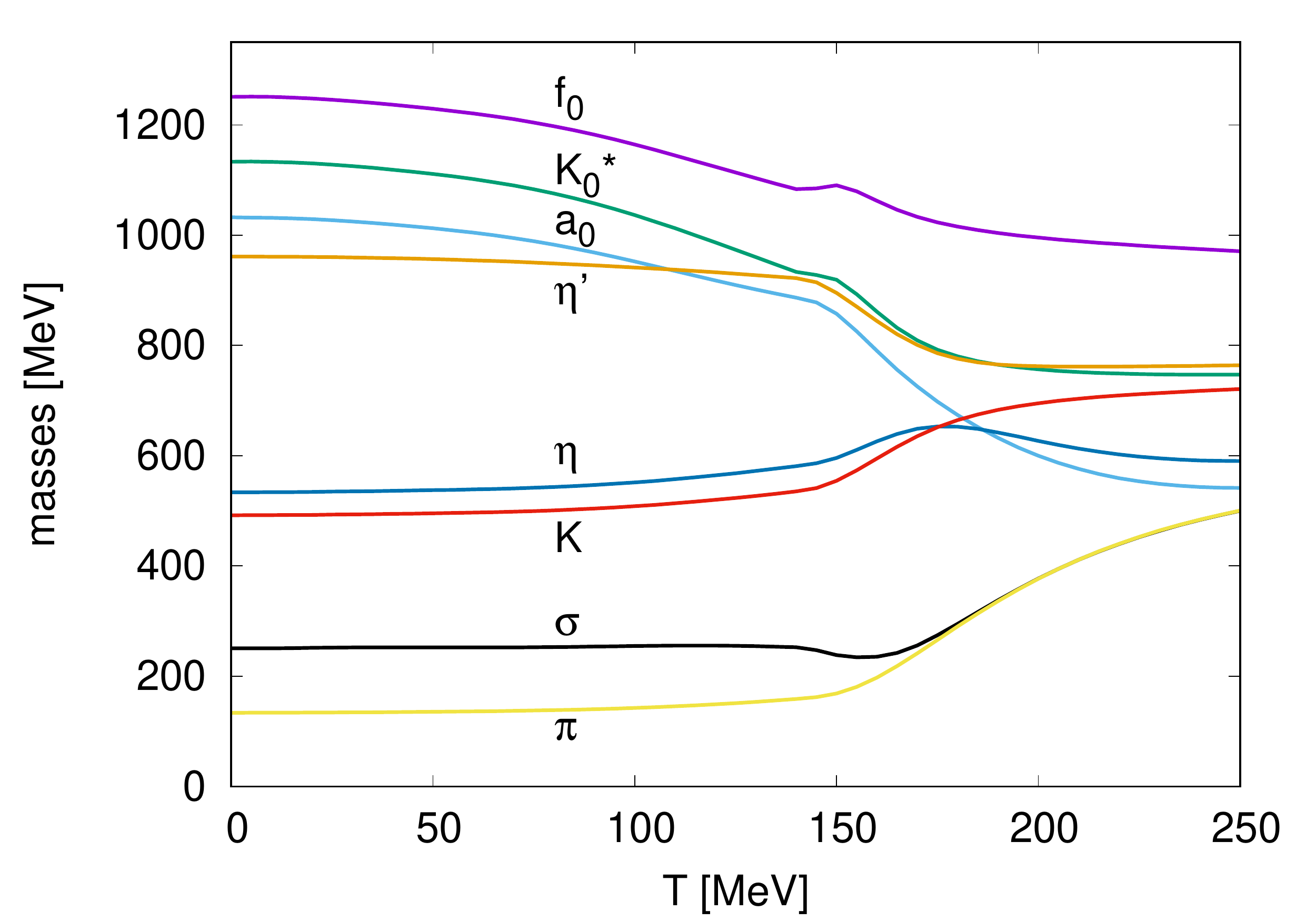}
\caption{Thermal behavior of the spectrum with instanton corrections. Nonphysical small bumps in some of the masses are presumably caused by the singular behavior of the derivative of the approximate $a(T)$ function.}
\label{Fig:mass2}
\end{figure}  

Throughout the calculations, the average instanton size is set to $r \simeq 1/3 \fm$, and the $T_0$ parameter has to be $T_0 \approx 143 \MeV$. In Fig. \ref{Fig:anom}, we show the absolute value of the dressed anomaly parameter in the minimum point of the effective potential, $|A|(\rho=\rho_{\min})$, as a function of the temperature, for all four possibilities. As expected, if no instanton contribution is present, then mesonic fluctuations strengthen the anomaly as the temperature rises. If the semiclassical tunelling amplitude is applied to the whole temperature range, then the anomaly coefficient monotonically decreases with the temperature, but in this case, the critical temperature comes out too small. If the instanton effects are taken into account only beyond $T_c$, then we see a momentary strengthening of the anomaly before it starts to drop. If we correct this scenario so that the critical temperature retains its physical value, we still get a very similar curve. The conclusion is that mesonic fluctuations can increase the anomaly up to about $\sim {\cal O}(10\%)$, before the instanton effects turn out to be more dominant and recover the $U_A(1)$ symmetry. As a result, it is seen that even at $\sim 1.5 T_c$ the anomaly is still visible, and the dressed $A$ parameter retains about $\sim 20\%$ of its $T=0$ value.  

We would like to draw attention to the fact that a similar evolution of the effective KMT coupling with respect to the temperature was also reported within the three flavor chiral Nambu--Jona-Lasinio model in Ref. \cite{fukushima00}. In the aforementioned study the authors investigated how the effective KMT coupling should be chosen as a function of $T$ so that lattice results for $\chi_{\top}$ can be reproduced the most accurately. Note that, even though $\chi_{\top}$ may monotonically decrease with $T$, the effective KMT coupling might not follow such behavior. As also discussed in Ref. \cite{fukushima00}, we also believe that $\chi_{\top}$ might not be the most appropriate quantity to measure the $U_A(1)$ breaking, as it entangles with the chiral condensates and can decrease, while the anomaly (the effective KMT coupling, to be precise) is still visible.

Along the same line of thinking, one may also attempt to characterize the increasing suppression of the anomaly via a dimensionless parameter that shows the level of $U_A(1)$ breaking in the effective potential. Since the KMT determinant is $\Delta=v_{\ns}^2v_{\s}/2\sqrt{2}$ in the physical background, we define $r_A$ as
\bea
r_A = \frac{[v_{\ns}^2v_{\s}A(\rho=\rho_{\min})]|_T}{[v_{\ns}^2v_{\s}A(\rho=\rho_{\min})]|_{T=0}},
\eea
see its temperature dependence in Fig. \ref{Fig:anom2}.  This quantity shows to what extent the actual contribution of the $U_A(1)$ breaking term of the potential at some temperature $T$ compares to its own value at $T=0$. Note that $r_A$, similarly to $\chi_{\top}$, entangles with the condensates, and as such, on top of the anomaly evolution, it also measures to what extent chiral symmetry is broken. As a result, $r_A$ should also be considered a less adequate quantity for characterizing purely the $U_A(1)$ breaking.

Finally,  as already discussed in the Introduction, we mention that the $a_0$-$\pi$ mass difference can be seen as a better indicator of the $U_A(1)$ restoration. Using the mass matrices (\ref{Eq:M2s}) and (\ref{Eq:M2pi}), we get $m_{a_0}^2 - m_{\pi}^2 = -\sqrt2 A(\rho_{\min}) v_{\s} + C(\rho_{\min})v^2_{\ns}$. That is to say, if the nonstrange condensate has significantly evaporated, the mass difference depends solely on the anomaly, assuming that the strange condensate does not change much with $T$. In other words, if beyond $T_c$ the aforementioned mass difference does not vanish, then the anomaly is still visible. In Fig. \ref{Fig:mass2}, we show the instanton corrected thermal behavior of the mass spectrum, realized in the most realistic scenario iii). The $a_0$-$\pi$ masses tend to get closer with $T$, but their difference shows that the anomaly does carry significance up to around $\sim 1.5 T_c$. One can check explicitly via the numerics that around this temperature the term proportional to $C(\rho_{\min})$ is indeed negligible, and therefore the $a_0$-$\pi$ mass difference is controlled by the anomaly alone.

\section{Discussion}

One of the main points of the paper is that in effective meson models that describe chiral symmetry restoration at finite temperature, perturbative treatments are not satisfactory. Couplings that receive field dependence (which can also be thought of as resummation of nonrenormalizable operators) through fluctuations do not even approximately behave as constants when mesonic fluctuations are integrated out; see, e.g., the solutions of $C_{k=0}$ and $A_{k=0}$ as the function of the chiral condensates in Figs. \ref{Fig:C} and \ref{Fig:A}. This raises doubts on treatments that perform perturbative corrections on vertices of the effective potential and points in the direction that resummation is a necessity. The functional renormalization group, which in effect was designed for resumming the field dependence of zero momentum vertices in a comparatively simple manner, is shown to be one of the most effective tools to obtain such nonperturbative results.

Another important result of the study is that the behavior of the absolute value of  dressed KMT determinant coupling can get larger when the temperature increases toward $T_c$. There are two distinct sources of such strengthening. On the one hand, the fully dressed, fluctuation corrected, field-dependent $A(\rho)$ anomaly coefficient function becomes explicitly temperature dependent, and on the other hand, since the minimum point of the effective potential corresponding to the chiral combination $\rho$ gets smaller as the temperature increases,  $A(\rho)$ has to be evaluated at different points so that an effective interaction can be defined. At growing temperatures, before instanton effects would recover $U_A(1)$ symmetry, the KMT coupling can acquire a qualitatively visible $\sim 10\%$ relative growth. This is in line with earlier expectations \cite{fejos16}; however, the effect appears to become more moderate from a quantitative point of view.

As analyzed in Sec. II, the applied chiral invariant expansion technique could be improved regarding the KMT term, since in the $U_V(3)$ vacuum it does not vanish. A more appropriate treatment would be to promote $U(\rho) \rightarrow U(\rho, \Delta)$, and solve its own flow equation in a two-dimensional grid. Together with field-dependent wave function renormalization, it might lead to improved scalar spectra. Of course, it is much more challenging from a numerical point of view; therefore, one might be interested in investigating the $U(\rho,\Delta) \approx U(\rho)+{\cal A}(\Delta)$ approximation, which leads to one-dimensional equations but would still resum all powers of $\Delta$ in the effective potential. As analyzed in Ref. \cite{pisarski20},  these terms can be associated with instanton configurations of higher topological charges. These directions represent active studies that will be reported elsewhere.

\section*{Acknowledgments}

This research was supported by the Hungarian National Research, Development and Innovation Fund under Projects No. PD127982 and K123815. G.F. was also supported by the János Bolyai Research Scholarship of the Hungarian Academy of Sciences and by the ÚNKP-21-5 New National Excellence Program of the Ministry for Innovation and Technology from the source of the National Research, Development and Innovation Fund.

\makeatletter
\@addtoreset{equation}{section}
\makeatother 
\renewcommand{\theequation}{A\arabic{equation}} 

\appendix

\section{Mass matrices}
The second derivative matrix of $V_k$, defined in (\ref{Eq:pot}), with $D_k=0$ can be written as
\bea
V''=\begin{pmatrix}
M^2_{s,k} & M^2_{s\pi,k} \\
M^2_{\pi s,k}  & M^2_{\pi,k} \\
\end{pmatrix},
\eea
where
\bea
\label{Eq:M2s}
(M_{s,k}^2)_{ij}&=& \delta_{ij}\Big(U_k'(\rho)+ C_k'(\rho)\tau+A_k'(\rho)\Delta \Big)\nonumber\\
&+&\frac{\partial^2 \tau}{\partial s_i\partial s_j}C_k(\rho)+\frac{\partial^2 \Delta}{\partial s_i\partial s_j}A_k(\rho)\nonumber\\
&+&\frac{\partial \rho}{\partial s_i}\frac{\partial \rho}{\partial s_j}\Big(U_k''(\rho)+C_k''(\rho)\tau+A_k''(\rho)\Delta\Big)\nonumber\\
&+&\Big(\frac{\partial \rho}{\partial s_i}\frac{\partial \tau}{\partial s_j}+\frac{\partial \rho}{\partial s_j}\frac{\partial \tau}{\partial s_i}\Big)C_k'(\rho)\nonumber\\
&+&\Big(\frac{\partial \rho}{\partial s_i}\frac{\partial \Delta}{\partial s_j}+\frac{\partial \rho}{\partial s_j}\frac{\partial \Delta}{\partial s_i}\Big)A_k'(\rho),
\eea
\bea
\label{Eq:M2pi}
(M_{\pi,k}^2)_{ij}&=& \delta_{ij}\Big(U_k'(\rho)+ C_k'(\rho)\tau+A_k'(\rho)\Delta \Big)\nonumber\\
&+&\frac{\partial^2 \tau}{\partial \pi_i\partial \pi_j}C_k(\rho)+\frac{\partial^2 \Delta}{\partial \pi_i\partial \pi_j}A_k(\rho)\nonumber\\
&+&\frac{\partial \rho}{\partial \pi_i}\frac{\partial \rho}{\partial \pi_j}\Big(U_k''(\rho)+C_k''(\rho)\tau+A_k''(\rho)\Delta\Big)\nonumber\\
&+&\Big(\frac{\partial \rho}{\partial \pi_i}\frac{\partial \tau}{\partial \pi_j}+\frac{\partial \rho}{\partial \pi_j}\frac{\partial \tau}{\partial \pi_i}\Big)C_k'(\rho)\nonumber\\
&+&\Big(\frac{\partial \rho}{\partial \pi_i}\frac{\partial \Delta}{\partial \pi_j}+\frac{\partial \rho}{\partial \pi_j}\frac{\partial \Delta}{\partial \pi_i}\Big)A_k'(\rho),
\eea
\bea
(M_{s\pi,k}^2)_{ij}&=&\frac{\partial^2 \tau}{\partial s_i\partial \pi_j}C_k(\rho)+\frac{\partial^2 \Delta}{\partial s_i\partial \pi_j}A_k(\rho)\nonumber\\
&+&\frac{\partial \rho}{\partial s_i}\frac{\partial \rho}{\partial \pi_j}\Big(U_k''(\rho)+C_k''(\rho)\tau+A_k''(\rho)\Delta\Big)\nonumber\\
&+&\Big(\frac{\partial \rho}{\partial s_i}\frac{\partial \tau}{\partial \pi_j}+\frac{\partial \rho}{\partial \pi_j}\frac{\partial \tau}{\partial s_i}\Big)C_k'(\rho)\nonumber\\
&+&\Big(\frac{\partial \rho}{\partial s_i}\frac{\partial \Delta}{\partial \pi_j}+\frac{\partial \rho}{\partial \pi_j}\frac{\partial \Delta}{\partial s_i}\Big)A_k'(\rho).
\eea
These matrix elements need to be calculated in a suitable background, before inserting them into the rhs of the flow equation (\ref{Eq:flow}).

\renewcommand{\theequation}{B\arabic{equation}} 

\section{Matsubara sums}

We define two basic sums, from which all others that are needed can be derived via differentiation with respect to the $\{\alpha_i\}$ parameters. As before, $\omega_n=2\pi n T$ are bosonic Matsubara frequencies,
\bea
{\cal S}_0(\alpha_0,\alpha_2,\alpha_4) = T \sum_{n=-\infty}^{\infty} \frac{1}{\alpha_0+\alpha_2\omega_n^2+\alpha_4\omega_n^4}, \\
{\cal S}_2(\alpha_0,\alpha_2,\alpha_4) = T \sum_{n=-\infty}^{\infty} \frac{\omega_n^2}{\alpha_0+\alpha_2\omega_n^2+\alpha_4\omega_n^4}.
\eea
These summations can be performed explicitly, and one arrives at
\bea
{\cal S}_0(\alpha_0,\alpha_2,\alpha_4) &=&\nonumber\\
&&\hspace{-1cm} \frac{\sqrt{\alpha_2-\sqrt{\alpha_2^2-4\alpha_0\alpha_4}}(\alpha_2+\sqrt{\alpha_2^2-4\alpha_0\alpha_4})}{4\sqrt{2 \alpha_4} \alpha_0 \sqrt{\alpha_2^2-4\alpha_0\alpha_4}}\nonumber\\
&&\hspace{-1cm}\times\coth \Bigg(\frac{\sqrt{\alpha_2-\sqrt{\alpha_2^2-4\alpha_0\alpha_4}}}{2\sqrt2 \sqrt{\alpha_4} T} \Bigg)\nonumber\\
&&\hspace{-1cm}+\frac{\sqrt{\alpha_2+\sqrt{\alpha_2^2-4\alpha_0\alpha_4}}(-\alpha_2+\sqrt{\alpha_2^2-4\alpha_0\alpha_4})}{4\sqrt{2 \alpha_4} \alpha_0 \sqrt{\alpha_2^2-4\alpha_0\alpha_4}}\nonumber\\
&&\hspace{-1cm}\times\coth \Bigg(\frac{\sqrt{\alpha_2+\sqrt{\alpha_2^2-4\alpha_0\alpha_4}}}{2\sqrt2 \sqrt{\alpha_4} T} \Bigg),
\eea
\bea
{\cal S}_2(\alpha_0,\alpha_2,\alpha_4) &=&-\frac{\sqrt{\alpha_2-\sqrt{\alpha_2^2-4\alpha_0\alpha_4}}}{4\sqrt{2 \alpha_4} \sqrt{\alpha_2^2-4\alpha_0\alpha_4}}\nonumber\\
&&\hspace{-1cm}\times\coth \Bigg(\frac{\sqrt{\alpha_2-\sqrt{\alpha_2^2-4\alpha_0\alpha_4}}}{2\sqrt2 \sqrt{\alpha_4} T} \Bigg)\nonumber\\
&&\hspace{-1cm}+\frac{\sqrt{\alpha_2+\sqrt{\alpha_2^2-4\alpha_0\alpha_4}}}{4\sqrt{2 \alpha_4} \sqrt{\alpha_2^2-4\alpha_0\alpha_4}}\nonumber\\
&&\hspace{-1cm}\times\coth \Bigg(\frac{\sqrt{\alpha_2+\sqrt{\alpha_2^2-4\alpha_0\alpha_4}}}{2\sqrt2 \sqrt{\alpha_4} T} \Bigg).
\eea

\end{document}